\documentclass[aps,prb,twocolumn,showpacs,superscriptaddress,10pt,nofootinbib]{revtex4}
 \usepackage[dvips]{graphicx}
 \usepackage{graphicx}
 \usepackage{gensymb}
 \usepackage{amsmath}
\usepackage{mathrsfs}
\usepackage{ulem}
\usepackage[mathscr]{euscript}
\usepackage{fancyhdr}
\usepackage{setspace}
\usepackage[usenames,dvipsnames]{color}

\begin{document}

\title{Novel Two-dimensional Carbon Allotrope with Strong Electronic Anisotropy}

\author{Cong Su}
\affiliation{International Center for Quantum Materials, Peking University, Beijing 100871, China}
\affiliation{Yuanpei College, Peking University, Beijing 100871, China}

\author{Hua Jiang}
\email{jianghuaphy@suda.edu.cn}
\affiliation{International Center for Quantum Materials, Peking University, Beijing 100871, China}
\affiliation{Department of Physics, Soochow University, Suzhou 215006, China}

\author{Ji Feng}
\email{jfeng11@pku.edu.cn}
\affiliation{International Center for Quantum Materials, Peking University, Beijing 100871, China}

\begin{abstract}
  Two novel two-dimensional carbon allotropes comprised of octagons and pentagons are proposed based on the first-principle calculations. The two carbon allotropes, named OPG-L and OPG-Z, are found to have distinct properties. OPG-L is metallic, while OPG-Z is a gapless semimetal. Remarkably, OPG-Z exhibits pronounced electronic anisotropy with highly anisotropic Dirac points at the Fermi level. A tight-binding model is suggested to describe the low-energy quasiparticles, which clarifies the origin of the anisotropic Dirac points. The electronic anisotropy of OPG-Z is expected to have interesting potential applications in electronic devices.
\end{abstract}


\maketitle

\section{introduction}

There have been growing interests in exploring new structures of the two-dimensional (2D) carbon in recent years. This is primarily stimulated by extensive investigations on the intriguing properties of graphene \cite{Graphene}, an atomically thin semimetal that harbors Dirac fermions at a pair of inequivalent valleys in the \textbf{k}-space \cite{Graphene_eletronic}. Among others, graphyne, graphdiyne, graphane, the $sp^2$-like carbon layer with five-, six- and seven-membered rings, the 2D amorphous carbon with four-membered rings, the planar carbon pentaheptite, the 2D carbon semiconductor with patterned defects, several carbon networks, octagraphene and T graphene \cite{graphyne,graphdiyne,graphane,5-6-7-rings,amorphous-2D-C,pentaheptite,caron-semiconductor,Graphene_allotropes,octagraphene,T_graphene} have been studied theoretically. In a clever synthetic attempt, graphdiyne has been successfully prepared experimentally \cite{graphdiyne}. One-dimensional (1D) topological defect containing octagonal and pentagonal $sp^{2}$-hybridized carbon rings embedded in a perfect graphene has been studied by first-principles approach \cite{ld_2} and produced experimentally \cite{line-defect}. The result is a line defect that mimics the 1D metallic wire, which has potential application in all-carbon valleytronics \cite{valley_filter, quantum_channel}. Consequently, one naturally suspects that the other 2D metastable carbon allotropes with intriguing properties may be prepared, in particular, comprised of five- and eight-membered carbon rings as inspired by the line defect in graphene.

\begin{figure}[!b]
\begin{center}
    \includegraphics[width=0.95\linewidth,clip]{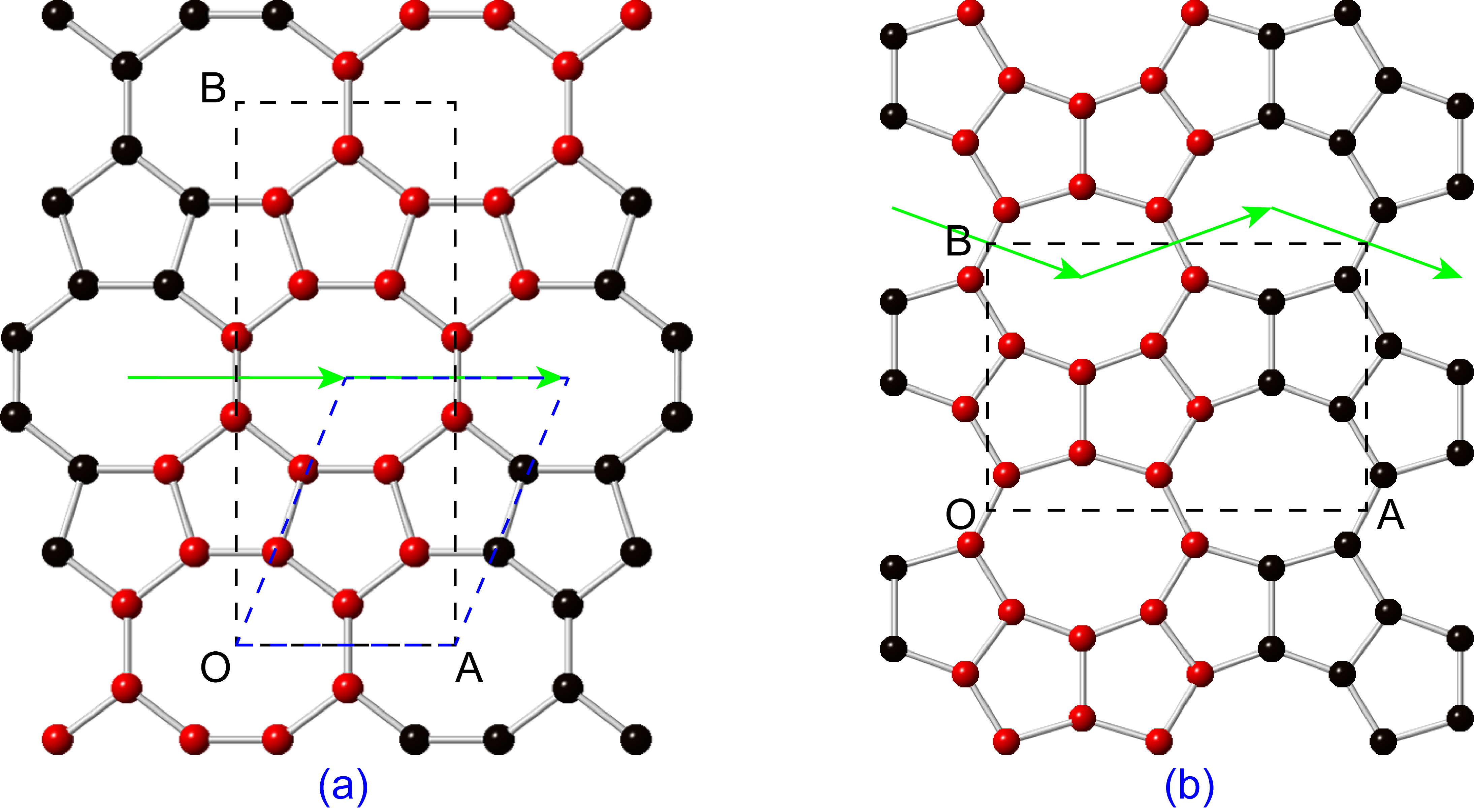}\\
\end{center}
\caption{The structures of (a) OPG-L and (b) OPG-Z. The black dashed frames are the orthogonal unit cells of OPG-L and OPG-Z, where OA and OB are lattice vectors. These two structures can be tiled by the red-colored 558 structure by copying it along the green arrows. The primitive cell of OPG-L is shown in blue dashed frame while the primitive cell of OPG-Z is the same as the crystal cell. Here we give the structural information of unit cells from the DFT calculations. OPG-L: space group $Cmmm$, OA = 3.68 \r{A}, OB = 9.12 \r{A}, two atoms in the asymmetric unit cell, (0, 0.42) and (0.69, 0.33); OPG-Z: space group $Pmam$, OA = 6.90 \r{A}, OB = 4.87 \r{A}, four atoms in asymmetric unit cell, (0.45,0.87), (0.56,0.62), (0.25,0.48), (0.25,0.78).} \label{fig:1}
\end{figure}

\section{Structures of OP-graphenes}

Apparently, viewing the line defects as structural motifs whose tiling covers the 2D plane is a sensible pathway for discovering new stable 2D carbon allotropes. Indeed, such tilings are geometrically viable. In this work, a 2D carbon allotrope is suggested to have an intrinsic strong electronic anisotropy, without the need for an external field \cite{anisotropic}. By using the first-principles calculations, we propose two novel energetically competitive, kinetically stable 2D carbon allotropes. They can be viewed as 2D tessellations of octagons and pentagons, called OP graphene-L (OPG-L) and OPG-Z, as shown in Figs. \ref{fig:1}(a) and (b), respectively. The structure of OPG-L can be viewed as juxtaposing the five-five-eight-membered rings (558) ribbon (indicated by the red atoms in Fig. 1) along a straight line path, while the OPG-Z along a zigzag path, as indicated by the green arrows in Figs. \ref{fig:1} (a) and (b). It is worth noting that this 558 ribbon occurs experimentally as a topological line defect of graphene \cite{line-defect}.  We show computationally that the OPG-L is a metal and OPG-Z is a gapless semimetal. Analysis of the electronic structures reveal that OPG-Z displays a strong electronic anisotropy, with anisotropic Dirac points near the Fermi level. These novel 2D carbon structures, with the proposed electronic properties, may be useful for novel electronic applications, in particular in all-carbon electronics \cite{dopants}.

\section{Computational method}

Our calculations are based on the density functional theory (DFT) within the generalized gradient approximation (GGA), in the form of Perdew-Burke-Ernzerhof's exchange-correlation functional \cite{PBE}. All the calculations are performed using the Vienna Ab-initio Simulation Package (VASP) \cite{VASP}. Periodic boundary conditions were employed and vacuum slabs of 10 {\AA} were used to isolate the replicas of OPG layers. Geometrical optimizations are performed until the Hellmann-Feynman forces on the ions are less than $1.0 \times 10^{-4}$ eV/{\AA}. The plane-wave basis is used, with a cut-off of 700 eV that converges the total energy to 1 meV/atom. The Brillouin zone is sampled using $9 \times 9 \times1$ Monkhorst-Pack \textbf{k}-point scheme \cite{MP}. The phonon spectra are calculated using the finite-displacement method in a $3 \times 3 \times 1$ supercell \cite{phonopy,PLK_method}.

\begin{figure}[!b]
\begin{center}
    \includegraphics[width=0.95\linewidth,clip]{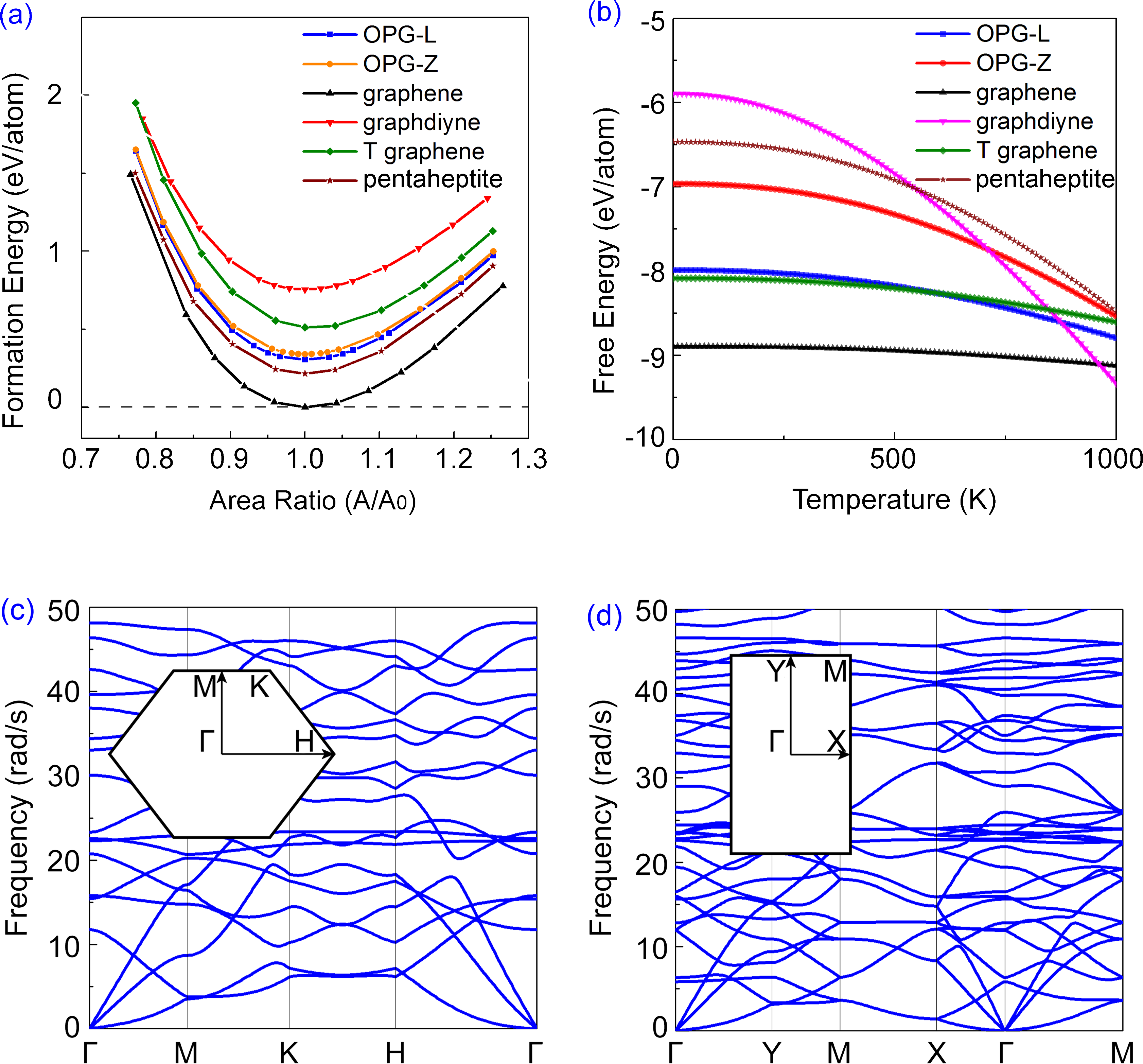}\\
\end{center}
\caption{(a) The formation energies of OPG-L, OPG-Z, graphdiyne, T graphene and pentaheptite as a function of area ratio in comparison to graphene, where $A_0$ is the optimized lattice area. The formation energy of graphene is set to 0. (b) The Helmholtz free energy as a function of temperature for the above-mentioned 2D carbon allotropes. The phonon spectra of (c) OPG-L and (d) OPG-Z. Inset: The first Brillouin zones and high symmetry points of OPG-L and OPG-Z. The high symmetry points are: $\Gamma$(0,0), M(0,0.5), K(0.419,0.709), H(0.581,0.291) in OPG-L, and $\Gamma$(0,0), X(0.5,0), M(0.5,0.5), Y(0,0.5) in OPG-Z (fractional coordinates in reciprocal space).}   \label{fig:2}
\end{figure}

\section{Stabilities of OP-Graphenes}

To gauge the stability of the proposed carbon structures, we calculated the formation energies, at $T = 0$ K within the static lattice approximation, of OPG-L, OPG-Z and four other typical 2D carbon allotropes, namely graphene, graphdiyne \cite{graphdiyne}, T graphene \cite{T_graphene} and pentaheptite \cite{pentaheptite} for comparison [Fig. \ref{fig:2}(a)]. The formation energy is defined with respect to the free-standing graphene. Among all the structures, graphene is the most stable energetically as expected \cite{T_carbon}. It is found that OPG-L and OPG-Z have fairly close formation energies, 0.31 eV/atom and 0.34 eV/atom respectively. Therefore, OPG-L and OPG-Z are energetically metastable compared to graphene and pentaheptite (whose formation energy is 0.21 eV/atom), though much stabler than previously proposed T graphene \cite{T_graphene} (0.52 eV/atom) and graphdiyne \cite{graphdiyne} (0.76 eV/atom). It should be noted that the successful synthesis of graphdiyne in a previous work \cite{graphdiyne} implies the realization of OPG-L and OPG-Z is not unlikely.

We further estimated the Helmholtz free energy as a function of temperature $T$. The Helmholtz free energy $A(T)$ is approximated as:
\begin{equation}
A(T)\approx A_{ph}(T)+E(0)
\end{equation}
where $A_{ph}(T)$ is the vibrational free energy within the Born-Oppenheimer and quasiharmonic approximation, and $E(0)$ is the total static-lattice energy at 0 K. The finite-temperature Fermi-Dirac distribution of electronic level occupation and the electron-phonon coupling are neglected. The vibrational free energy is calculated by
\begin{equation}
A_{ph}(T)=\frac{1}{2}\sum_{\textbf{q},s}\hbar\omega(\textbf{q},s)+k_BT\sum_{\textbf{q},s}\ln[1-exp(-\frac{\hbar\omega(\textbf{q},s)}{k_BT})]
\end{equation}
where \textbf{q} stands for the wavevector, $s$ the branch index, $\omega$ the frequency at 0 K, $k_B$ and $\hbar$ the Boltzmann and Planck's constants. As a result, we find that the free energy of OPG-L and OPG-Z [Fig. \ref{fig:2}(b)] falls rapidly with increasing temperature. At low temperature, OPGs are thermodynamically stabler than graphdiyne and pentaheptite. The phonon spectra of OPG-L and OPG-Z are also calculated [see Fig. \ref{fig:2}(c) and (d)] and no imaginary phonon modes are found, confirming again the kinetic stability of these two carbon sheets. In addition, the chemical stability of these two structures are also examined. We put dioxygen and dihydrogen molecules close to OPG-L and OPG-Z. After geometric relaxations, these molecules are all repelled by the carbon sheet. Therefore, no spontaneous chemical reaction is expected between these structures and oxygen or hydrogen molecules, suggesting their redox stability in atmosphere.

\begin{figure}[!b]
\begin{center}
    \includegraphics[width=0.95\linewidth,clip]{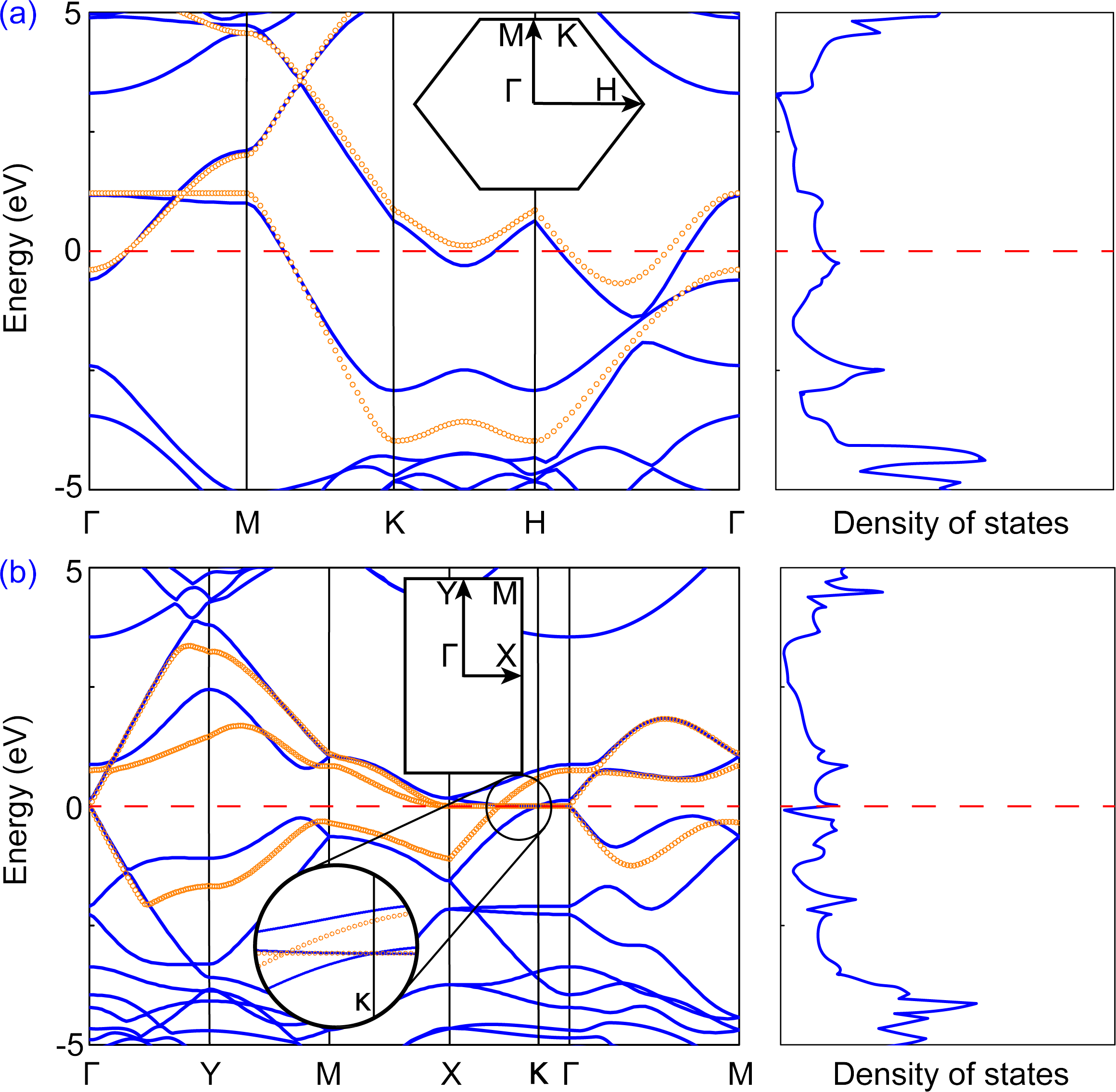}\\
\end{center}
\caption{Electronic band structures and density of states for (a) OPG-L and (b) OPG-Z respectively. The inset figures are locations of high symmetry points. The red dashed lines represent Fermi levels, which are set to 0 eV. Blue lines are the results of DFT calculations, while orange dotted lines are the results of tight-binding model.}   \label{fig:3}
\end{figure}

\begin{figure*}[!t]
\begin{center}
    \includegraphics[width=0.9\linewidth,clip]{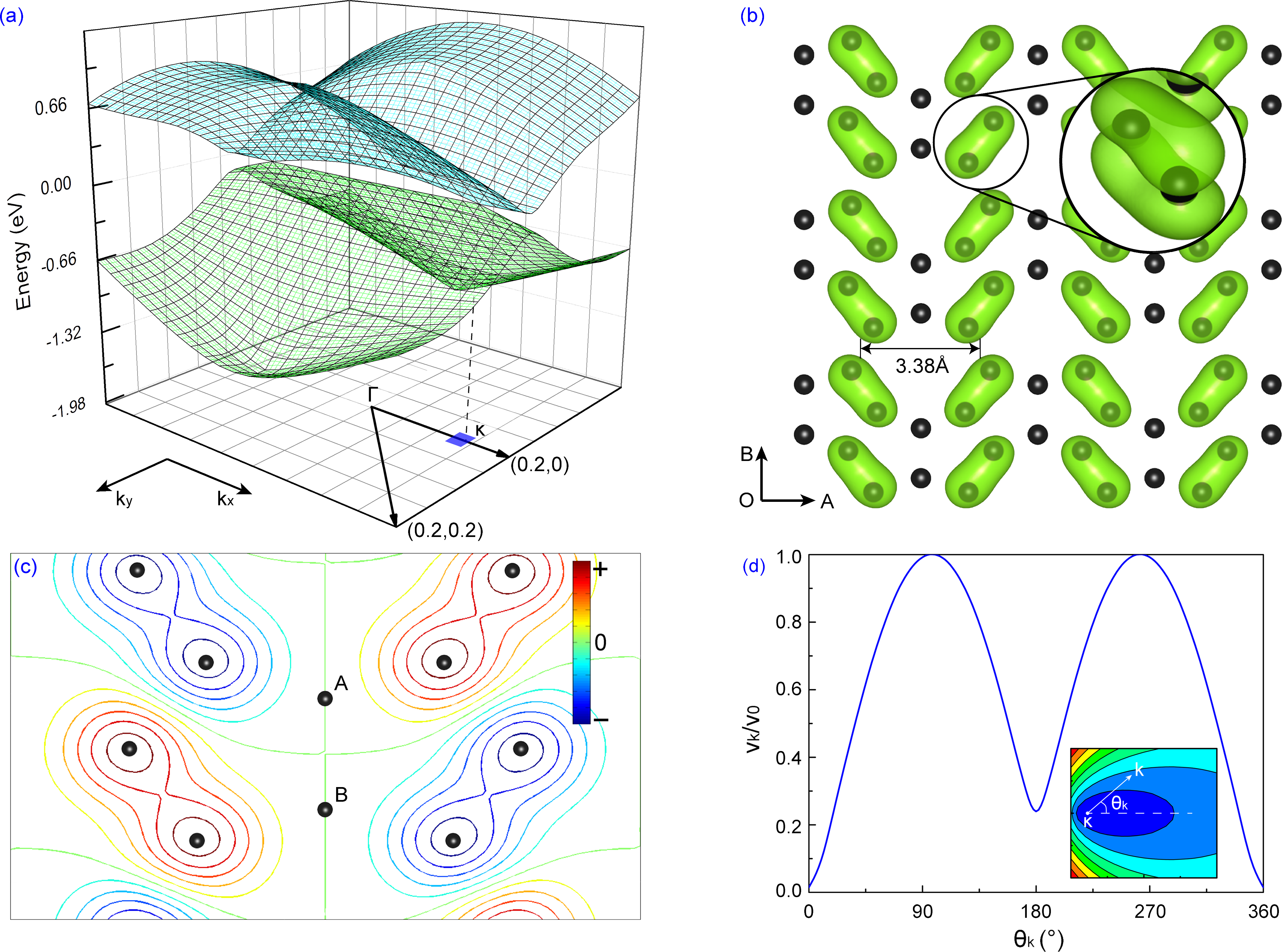}\\
\end{center}
\caption{(a) The blow-up band structures of Kohn-Sham quasiparticles of OPG-Z near the Fermi level around $\Gamma$ point. (b) The partial charge density (PCD) at $\kappa$ point near Fermi level, calculated by DFT.  (c) The wave function in a unit cell at the same point, calculated by the tight-binding model [Eq.(1)]. (d) The component of group velocity parallel to the \textbf{k} vector ($v_k$) measured from the $\kappa$ point in units of the Fermi velocity along $k_y$ ($v_0$) versus the angle ($\theta_k$) of the \textbf{k} vector. Inset:  The isosurface of the band structure of OPG-Z within the range of $k_x\in [0.139,0.141]$ and $k_y\in [-0.0001,0.0001]$. The black dots in (b) and (c) are carbon atoms.}   \label{fig:4}
\end{figure*}

\begin{figure*}[!tb]
\begin{center}
    \includegraphics[width=0.8\linewidth,clip]{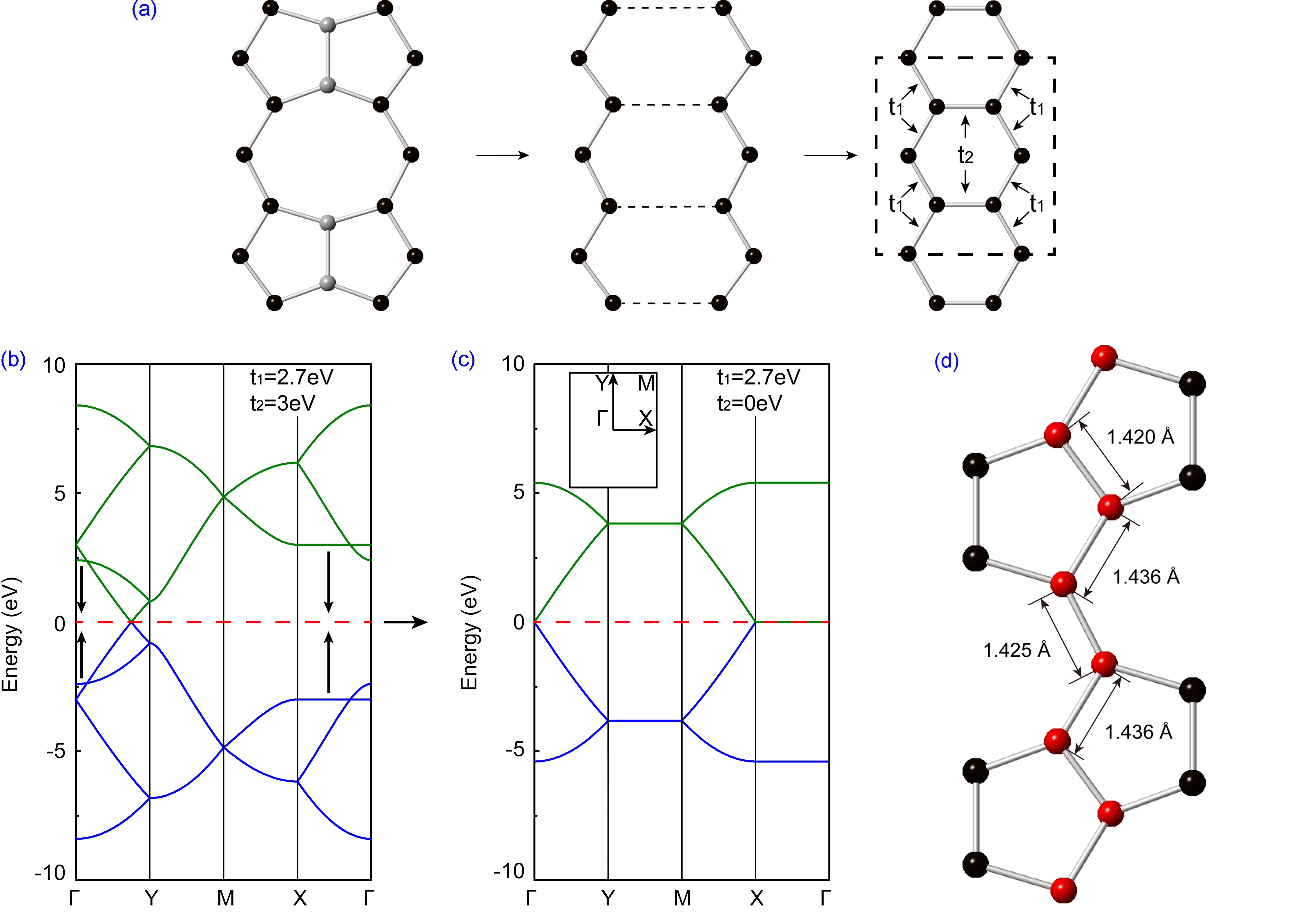}\\
\end{center}
\caption{(a) OPG-Z as a distorted honeycomb lattice. Gray atoms are the atoms with zero wavefunction amplitude around Fermi level in OPG-Z. The supercell of graphene consists of eight atoms with its TB hopping matrix elements $t_1$ and $t_2$ shown. (b) The tight-binding band structure of graphene in the supercell shown in (a), with $t_2=3$ eV. (c) The TB band structure of graphene of the same primitive cell, with $t_2=0$ eV. Inset: high symmetry points of the graphene supercell. (d) The different bond lengths of the zigzag ribbon in OPG-Z.}   \label{fig:5}
\end{figure*}

\section{electronic structures and possible realization}
\subsection{Energy Bands}

The Kohn-Sham electronic band structures and density of states (DOS) of OPG-L and -Z are shown in Fig. \ref{fig:3}. As we can see, OPG-L is a metal and OPG-Z is a gapless semimetal. In Fig. \ref{fig:3}(b), $\Gamma=$(0,0) point (fractional coordinates with respect to the reciprocal unit cell; we follow the same notation hereafter) near the Fermi level (FL) is dominated by linear dispersion, resembling the Dirac fermions. A closer inspection reveals that OPG-Z has a wedge-shaped conduction band and valence band near the FL where the Dirac points emerge at $\kappa = (0.139,0)$ and $\kappa' = (-0.139,0)$, as shown in Fig. \ref{fig:4}(a). The Fermi velocity along $k_y$ at these Dirac points are estimated to be $2.2 \times 10^{5}$ m/s, relatively high but an order of magnitude smaller than in pristine graphene. The Kohn-Sham quasiparticle energy isosurfaces of conduction band of OPG-Z near $\kappa$ point are very elongated ellipses with an eccentricity near unity [inset of Fig. \ref{fig:4}(d)]. In addition, the Fermi velocities along different directions measured from $\kappa$ point are also calculated [Fig. 4(d)], which shows that the group velocity is redueced to $3.2 \times 10^{3}$ m/s in the $k_x$ direction. Therefore, OPG-Z is semimetallic with highly anisotropic Dirac fermions \cite{anisotropic}.

\subsection{Tight-Binding Approximation}
To further understand the DFT calculated results, we construct a tight-binding model of OPG-Z by allowing for the electrons hopping only to nearest neighbors. The Hamiltonian can be written as
\begin{equation}  \label{eq:1}
H=-\sum_{\langle i,j\rangle\sigma}t_{ij}(a_{i\sigma}^\dag a_{j\sigma} + h.c.) + \varepsilon_0\sum_{i,\sigma} a_{i\sigma}^\dag a_{i\sigma},
\end{equation}
where $a_{i\sigma}^\dag$ and $a_{i\sigma}$ are, respectively, the electronic creation and annihilation operators of the carbon valence $p_z$ orbital with spin $\sigma$ at site $i$; $\langle i,j\rangle$ stands for the nearest-neighbor pairs of atoms at sites $i$ and $j$; $\varepsilon_0$ is the on-site energy and $t_{ij}$ are the hopping matrix elements between site $i$ and site $j$, which are all taken to be 3.0 eV \cite{Review} except for $t_{AB}$ [1 eV, site $A$ and site $B$ are shown in Fig. \ref{fig:4}(c)]. Because all the atoms are carbons, the on-site energy $\varepsilon_0$ can be set to zero. We calculated the band structures (orange dotted lines in Fig. \ref{fig:3}), which are observed to match well the DFT calculations. The partial charge density at $\kappa$ point in the conduction band is also calculated by DFT, shown in Fig. \ref{fig:4}(b), which is in excellent agreement with the wavefunction calculated by the TB [Fig. \ref{fig:4}(c)]. Notice that the electron density is zero on atoms A and B in Fig. \ref{fig:4}(c), naturally splitting the whole plane into independent 1D electron channels along OB direction [Fig. 4(b)]. The continuous electron density along OB and the strong localization of electrons along OA bring about the anisotropic electronic conductivity in OPG-Z. In this case, a perfect 1D aligned electron transport is able to be realised in OPG-Z. Such stripe-like electron transport channels are as wide as 3.38 {\AA} [Fig. \ref{fig:4}(b)]. The anisotropic property of electronic transport in OPG-Z may give rise to some interesting implications in nano-electronics. For example, the 1D electronic channels in OPG-Z layer can be a replacement of the conducting wires in thin film transistor.

It is also of interest to analyze how the Dirac points and flat band emerge around FL in OPG-Z. We observe that the elimination of the nodal atoms [zero wavefunction amplitude near the Dirac point; c.f., Fig. \ref{fig:4}(b)] from OPG-Z leads a distorted graphene lattice [Fig. \ref{fig:5}(a)]. A TB model of graphene will be good to describe some of the OPG-Z's properties. The TB model has two hopping parameters, conforming to the prescribed lattice distortion, as shown in Fig. \ref{fig:5}(a). By expanding the primitive cell of graphene to an eight-atom supercell, two linear dispersion bands and flat bands at $E=\pm t_2$ are folded to the $\Gamma$ point in the first Brillouin zone [Fig. \ref{fig:5}(b)]. Upon setting $t_2$ to zero, these two sets of band are shifted to FL, creating one band with linear dispersion and one flat band at $\Gamma$ point [Fig. \ref{fig:5}(c)]. As the simple model demonstrates, the anisotropic property of OPG-Z is originated from the separation of zigzag ribbons in graphene. The spliting of Dirac point from $\Gamma$ point to $\kappa$ and $\kappa'$ points in OPG-Z arises from the further structural relaxation that leads to alternating bond lengths as indicated by the red atom chain in Fig. \ref{fig:5}(d).

\subsection{Possible Experimental Realization}
As its formation energy and free energy is lower than the existing carbon allotropes like graphdiyne\cite{graphdiyne}, OPG-Z might be experimentally obtained. Possible routes include nano-engineered synthesis \cite{Defect_exp,line_defect_TBMD,nano_eng}, epitaxial or chemical vapor deposition on a suitable substrate \cite{epitaxial,graphene,graphene_2} and chemically, acetylene scaffolding planar dicyclopenta pentalene \cite{scaffolding,four_five}.

\section{Summary}

In summary, with first-principle calculations we have shown that two new 2D carbon allotropes comprised of octagons and pentagons, named as OPG-L and OPG-Z can be energetically and kinetically viable. The energetic and kinetic stabilities of OPG-L and -Z are supported by calculating their zero temperature energies, approximate finite temperature free energy, and phonon band structure. Their electronic structures are calculated and analyzed in detail. Our results show that OPG-L and OPG-Z are even more favorable in energy than graphdiyne that was already synthesized experimentally. OPG-L is a metal while OPG-Z is a gapless semimetal. It is found that the electronic structure of OPG-Z is remarkably anisotropic, with a pair of anisotropic Dirac points very close to Fermi level. The wavefunction and partial charge density in the conduction band of OPG-Z at Dirac point, in combination with transparent TB model, explain the origin of the electronic anisotropy of OPG-Z. The computed stability and electronic structure argue for experimental synthesis of these two-dimensional carbon structures, which are expected to have potentially interesting applications.

\acknowledgments
We are grateful to Ran Duan and Dr. Haiwen Liu for useful discussions.
We thank the financial support from National Science Foundation of China (Grant No. 11174009, No. 91121004 and No. 11274032) and China 973 Project (Grant No. 2013CB921900).


\begin{thebibliography}{99}
\bibitem{Graphene} K. S. Novoselov, A. K.Geim, S. V. Morozov, D. Jiang, M. I. Katsnelson, I. V. Grigorieva, S. V. Dubonos, and A. A. Firsov, Nature (London) \textbf{438}, 197 (2005).
\bibitem{Graphene_eletronic} P.R. Wallace, Phys. Rev. \textbf{71}, 622 (1947).
\bibitem{graphyne} M. M. Haley, Pure Appl. Chem. \textbf{80}, 519 (2008); J. M. Kehoe \textit{et al.}, Org. Lett. \textbf{2}, 969 (2000).
\bibitem{graphdiyne} G. X. Li, Y. L. Li, H. B. Liu, Y. B. Guo, Y. J. Li and D. B. Zhu, Chem. Commun. \textbf{46}, 3256 (2010).
\bibitem{graphane} J. O. Sofo, A. S. Chaudhari, and G. D. Barber, Phys. Rev. B \textbf{75}, 153401 (2007); D. C. Elias \textit{et al.}, Science \textbf{323}, 610 (2009).
\bibitem{5-6-7-rings} H. Terrones, M. Terrones, E. Hern\'andez, N. Grobert, J-C. Charlier, and P. M. Ajayan, Phys. Rev. Lett. \textbf{84}, 1716 (2000).
\bibitem{amorphous-2D-C}J. Kotakoski, A.V. Krasheninnikov, U. Kaiser, and J. C. Meyer, Phys. Rev. Lett. \textbf{106}, 105505 (2011).
\bibitem{pentaheptite} V. H. Crespi, L. X. Benedict, M. L. Cohen, and S. G. Louie, Phys. Rev. B \textbf{53}, R13303 (1996).
\bibitem{caron-semiconductor} David J. Appelhans, Zhibin Lin, and Mark T. Lusk, Phys. Rev. B \textbf{82}, 073410 (2010).
\bibitem{Graphene_allotropes} A.N. Enyashin and A.L. Ivanovskii, Phys. Status Solidi B \textbf{248}, 1879 (2011).
\bibitem{octagraphene} X. L. Sheng, H. J. Cui, F. Ye, Q. B. Yan, Q. R. Zheng, and G. Su, J. Appl. Phys. \textbf{112}, 074315 (2012).
\bibitem{T_graphene} Y. Liu, G. Wang, Q. S. Huang, L. W. Guo and X. L. Chen, Phys. Rev. Lett. \textbf{108}, 225505 (2012).
\bibitem{ld_2} A. R. Botello-M\'{e}ndez, X. Declerck, M. Terrones and J.-C. Charlier, Nanoscale \textbf{3}, 2868 (2011).
\bibitem{line-defect} Jayeeta Lahiri, You Lin, Pinar Bozkurt, Ivan I. Oleynik, and Matthias Batzill, Nature Nanotech. \textbf{5}, 326 (2010).
\bibitem{valley_filter} D. Gunlycke, C. T. White, Phys. Rev. Lett. \textbf{106}, 136806 (2011).
\bibitem{quantum_channel} J. T. Song, H. W. Liu, H. Jiang, Q. -F. Sun and X. C. Xie, Phys. Rev. B \textbf{86}, 085437 (2012).
\bibitem{anisotropic} C. -H. Park, L. Yang, Y. -W. Son, M. L. Cohen, and S. G. Louie, Nature Phys. \textbf{4}, 213 (2008).
\bibitem{dopants} A. Das, S. Pisana, B. Chakraborty, S. Piscanec, S. K. Saha, U. V. Waghmare, K. S. Novoselov, H. R. Krishnamurthy, A. K. Geim, A. C. Ferrari and A. K. Sood, Nature Nanotech. \textbf{3}, 210 (2008).
\bibitem{PBE} J.P. Perdew, K. Burke and M. Ernzerhof, Phys. Rev. Lett. \textbf{77}, 3865 (1996).
\bibitem{VASP} G. Kresse and J. Furthm\"{u}ller, Phys. Rev. B \textbf{54}, 11169 (1996).
\bibitem{MP} H.J. Monkhorst and J.D. Pack, Phys. Rev. B \textbf{13}, 5188 (1976).
\bibitem{phonopy} A. Togo, F. Oba and I. Tanaka, Phys. Rev. B \textbf{78}, 134106 (2008).
\bibitem{PLK_method} K. Parlinski, Z. Q. Li, and Y. Kawazoe, Phys. Rev. Lett. \textbf{78}, 4063 (1997).
\bibitem{T_carbon} X. L. Sheng, Q. B. Yan, F. Ye, Q. R. Zheng and G. Su, Phys. Rev. Lett. \textbf{106}, 155703 (2011).
\bibitem{Review} A. H. Castro Neto, F. Guinea, N. M. R. Peres, K. S. Noselov and A. K. Geim, Rev. Mod. Phys. \textbf{81}, 109 (2009).
\bibitem{Defect_exp} A. Hashimoto, K. Suenaga, A. Gloter, K. Urita and S. Iijima, Nature (London) \textbf{430}, 870 (2004).
\bibitem{line_defect_TBMD} S. Okada, T. Kawai and K. Nakada, J. Phys. Soc. Jpn. \textbf{80}, 013709 (2011).
\bibitem{nano_eng} M. T. Lusk and L. D. Carr, Phys. Rev. Lett. \textbf{100}, 175503 (2008).
\bibitem{epitaxial} J. Hass, W.A. Heer and E.H. Conrad, J. Phys.: Condens. Mater. \textbf{20}, 323202 (2008).
\bibitem{graphene} X. S. Li, W. W. Cai, J. An, S. Kim, J. Nah, D. X. Y, R. Piner, A. Velamakanni, I. Jung, E. Tutuc, S. K. Banerjee, L. Colombo and R. S. Ruoff, Science \textbf{324}, 1312 (2009).
\bibitem{graphene_2} A. Reina, X. T. Jia, J. Ho, D. Nezich, H. Son, V. Bulovic, M. S. Dresselhaus and J. Kong, Nano Letters \textbf{9}, 30 (2009).
\bibitem{scaffolding} S. W. Slayden and J. F. Liebman, Chem. Rev. \textbf{101}, 1541 (2001); U. H. F. Bunz, Y. Rubin and Y. Tobe, Chem. Soc. Rev. \textbf{28}, 107 (1999).
\bibitem{four_five} H. Cao, S. G. V. Ornum, J. Deschamps, J. F.-Anderson, F. Laib and J. M. Cook, J. Am. Chem. Soc. \textbf{127}, 933 (2005).
\end{thebibliography}
\end{document}